\documentstyle[preprint,eqsecnum,graphics,aps]{revtex}

\begin{document}

\title{A gentle introduction to\\ the foundations of classical
  electrodynamics:\\ The meaning of the excitations $({\cal D}, {\cal H})$
  \\and the field strengths $(E, B)$}

\author{Friedrich W.\ Hehl\thanks{Email: hehl@thp.uni-koeln.de}\\ 
  Institute for Theoretical Physics\\ University of Cologne\\ 50923
  K\"oln, Germany \\ and \\Yuri N.\ 
  Obukhov\thanks{Email: general@elnet.msk.ru}\\ Department of
  Theoretical Physics\\ Moscow State University\\ 117234 Moscow,
  Russia}

\maketitle 

\begin{abstract} 
The axiomatic structure of the electromagnetic theory is outlined.
  We will base classical electrodynamics on (1) electric charge
  conservation, (2) the Lorentz force, (3) magnetic flux conservation,
  and (4) on the Maxwell-Lorentz spacetime relations.  This yields the
  Maxwell equations.  The consequences will be drawn, inter alia, for
  the interpretation and the dimension of the electric and magnetic
  fields. 
\end{abstract}

\newpage
\section{Introduction}

In Cologne, we teach a course on Theoretical Physics II
(electrodynamics) to students of physics in their fourth semester.
For several years, we have been using for that purpose the calculus of
exterior differential forms, see \cite{book,Zirnbauer}, because we
believe that this is the appropriate formalism: It is based on objects
which possess a clear operational interpretation, it elucidates the
fundamental structure of Maxwell's equations and their mutual
interrelationship, and it invites a 4-dimensional representation
appropriate for special {\em and} general relativity theory (i.e.,
including gravity, see \cite{OH,HO}).

Our experimental colleagues are somewhat skeptical; and not only them.
Therefore we were invited to give, within 90 minutes, a sort of 
popular survey of electrodynamics in exterior calculus to the members
of one of our experimental institutes (group of H. Micklitz). The
present article is a worked-out version of this talk. We believe that
it could also be useful for other universities.

Subsequent to the talk we had given, we found the highly interesting
and historically oriented article of Roche \cite{Roche} on ``$B$ and
$H$, the intensity vectors of magnetism\dots".  Therein, the
corresponding work of Bamberg and Sternberg \cite{Bamberg}, Bopp
\cite{Bopp}, Ingarden and Jamio{\l}kowski \cite{Ingarden}, Kovetz
\cite{Kovetz}, Post \cite{Post1}, Sommerfeld \cite{Sommerfeld}, and
Truesdell and Toupin \cite{Truesdell}, to drop just a few names, was
neglected yielding a picture of $H$ and $B$ which looks to us as being
not up of date; one should also compare in this context the letter of
Chambers \cite{Chambers} and the book of Roche \cite{Rochebook}, in
particular its Chapter 9.  Below we will suggest answers to some of
Roche's questions.

Moreover, ``...any system that gives $E$ and $B$ different units, when
they are related through a relativistic transformation, is on the far
side of sanity'' is an apodictic statement of Fitch \cite{Fitch}. In
the sequel, we will prove that we {\em are} on the far side of sanity:
The {\em absolute} dimension of $E$ turns out to be {\em magnetic
  flux/time} and that of $B$ {\em magnetic flux,} see
Sec.~\ref{strengths}.

According to the audience we want to address, we will skip all
mathematical details and take recourse to plausibility considerations.
In order to make the paper self-contained, we present though a brief
summary of exterior calculus in the Appendix. A good reference to the
mathematics underlying our presentation is the book of Frankel
\cite{Ted}, see also \cite{Bamberg} and \cite{Schouten}. For the
experimental side of our subject we refer to Bergmann-Schaefer
\cite{Raith}.

As a preview, let us collect essential information about the
electromagnetic field in Table I. The explanations will follow below.

\begin{center}
\noindent{\bf Table I.} The electromagnetic field\bigskip

\noindent
\begin{tabular}{|c|c|c|c|c|c|c|}
\hline 
Field & name   & math. & independent & related & reflec- &absolute \\
 & & object & components & to & tion & dimension \\
\hline 
${\cal D}$ & electric & odd & ${\cal D}_{23},{\cal D}_{31},
{\cal D}_{12}$  & area & $-{\cal D}$ & $q=$ electric \\
& excitation & 2-form & & & & charge \\
\hline
${\cal H}$ & magnetic & odd & ${\cal H}_1,{\cal H}_2,{\cal H}_3$ 
& line & $-{\cal H}$ & $q/t$\\
& excitation & 1-form & & && \\
\hline
$E$ & electric & even & $E_1,E_2,E_3$ & line & $E$ & $\Phi_0/t$\\
& field strength & 1-form & & & & \\
\hline
$B$ & magnetic & even & $B_{23},B_{31},B_{12}$ & 
area & $B$ & $\Phi_0=$ mag- \\
&  field strength & 2-form & & & & netic flux \\
\hline
\end{tabular}
\end{center}
\vskip3cm

It was Maxwell himself who advised us to be very careful in assigning
a certain physical quantity to a mathematical object. As it turns out,
the mathematical images of ${\cal D},{\cal H}, E, B$ are all different
{}from each other. This is well encoded in Schouten's images of the
electromagnetic field in Fig.1.

\begin{figure} 
\begin{center} 
\includegraphics{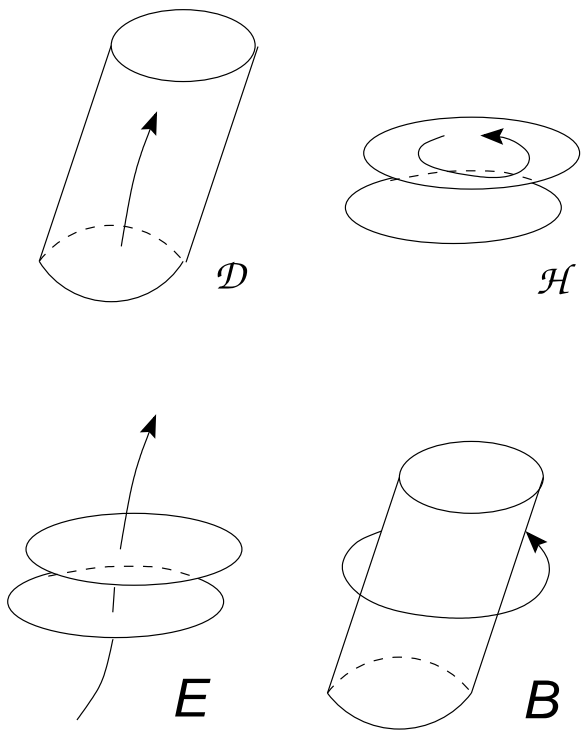} 
\caption{Schouten's images of the electromagnetic field, 
see [17] p.\ 133.} 
\end{center} 
\end{figure}

\section{Electric charge conservation}\label{chargecons}

The conservation of electric charge was already recognized as
fundamental law during the time of Franklin (around 1750) well before
Coulomb discovered his force law in 1785. Nowadays, when one can catch
single electrons and single protons and their antiparticles in traps
and can {\em count} them individually (see, e.g., Dehmelt
\cite{Dehmelt}, Paul \cite{Paul}, Devoret et al.\ \cite{Devoret1}, and
Lafarge et al.\ \cite{Devoret2}), we are more sure than ever that
electric charge conservation is a valid fundamental law of nature.
Therefore matter carries as a {\em primary quality} something called
electric charge which only occurs in positive or negative units of an
elementary charge $e$ (or, in the case of quarks, of $1/3$th of it)
and which can be counted. Thus it is justified to introduce the
physical dimension of charge $q$ as a new and independent concept.
Ideally one should measure a charge in units of $e/3$. However, for
practical reasons, the SI-unit C (coulomb) is used in laboratory
physics.

Let us start with the 3-dimensional Euclidean space in which we mark a
3-dimensional domain $V$. Hereafter, the local coordinates in this
space will be denoted by $x^a$ and the time by $t$, with the basis
vectors $e_a :=\partial_a$ and $a,b,\dots=1,2,3$, see Fig.2. The total
charge in the domain $V$ is given by the integral 
\begin{equation}
Q=\int\limits_V\rho\,,\label{Qrho} 
\end{equation} 
where the electric charge density $\rho$ is
the 3-form $\rho={\frac 1{3!}} \,\rho_{abc}\,dx^a\wedge dx^b\wedge
dx^c = \rho_{123}\,dx^1\wedge dx^2 \wedge dx^3$. Here summation is
understood over the indices $a,b,c$ and
$\rho_{abc}=-\rho_{bac}=\rho_{bca}=\dots$, i.e., the components
$\rho_{abc}$ of the charge density 3-form $\rho$ are antisymmetric
under the exchange of two indices, leaving only one independent
component $\rho_{123}$.  The wedge $\wedge$ denotes the
(anticommutative) exterior product of two forms, and $dx^1\wedge dx^2
\wedge dx^3$ represents the volume ``element''.  For our present
purpose it is enough to know, for more details see the Appendix, that
a 3-form (a $p$-form) is an object that, if integrated over a
3-dimensional ($p$-dimensional) domain, yields a scalar quantity, here
the charge $Q$.
\begin{figure} 
\begin{center} 
\includegraphics{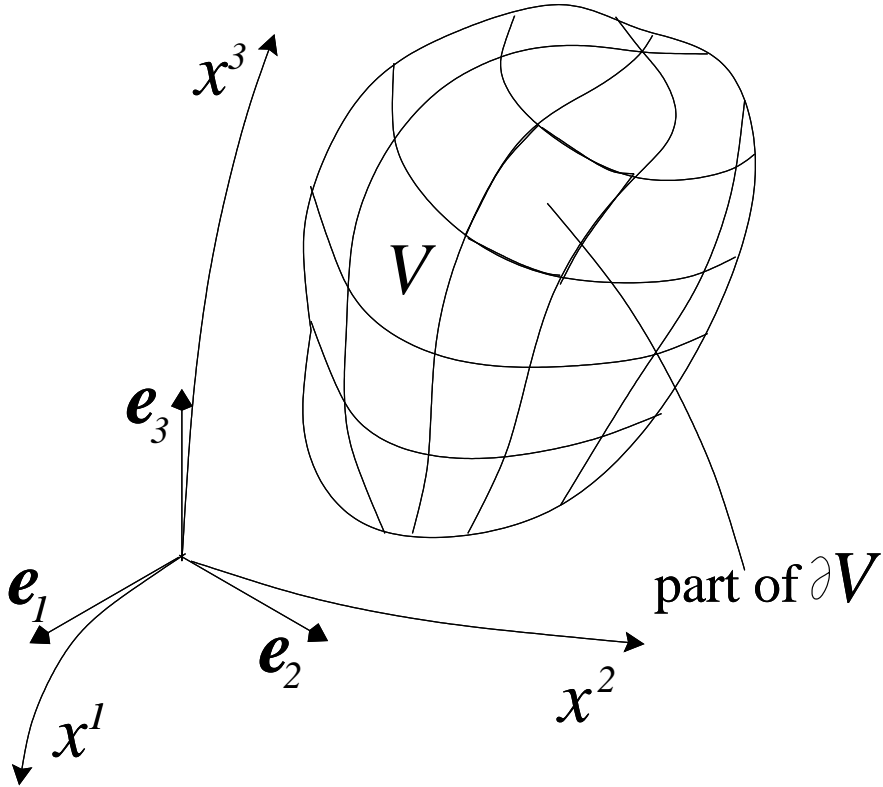} 
\caption{A volume $V$ with its boundary $\partial V$.} 
\end{center} 
\end{figure} 

The dimension of $Q$ is $[Q]=q$.  Since an integral (a summation after
all) cannot change the dimension, the dimension of the charge density
3-form and its components are, respectively, $[\rho]=q$, and
$[\rho_{abc}]=q/\ell^{3}$, with $\ell=$ length.

In general, the elementary charges are not at rest. The electric current 
$J$ flowing across a 2-dimensional surface $S$ is given by the integral 
\begin{equation}
J=\int\limits_S j\,,\label{Jj} 
\end{equation} 
see Fig.3. Accordingly, the electric current density $j$ turns out to
be a 2-form: $j={\frac 1 {2!}}j_{ab}\,dx^a \wedge dx^b
=j_{12}\,dx^1\wedge dx^2+j_{13}\,dx^1\wedge dx^3+j_{23}\, dx^2\wedge
dx^3$, with $j_{ab}=-j_{ba}$.  If $t=$ time, then the dimensions of
the current integral and the current 2-form and its components are
$[J]=[j]=q/t$ and $[j_{ab}]=q/(t\,\ell^{2})$, respectively.
\begin{figure} 
\begin{center} 
\includegraphics{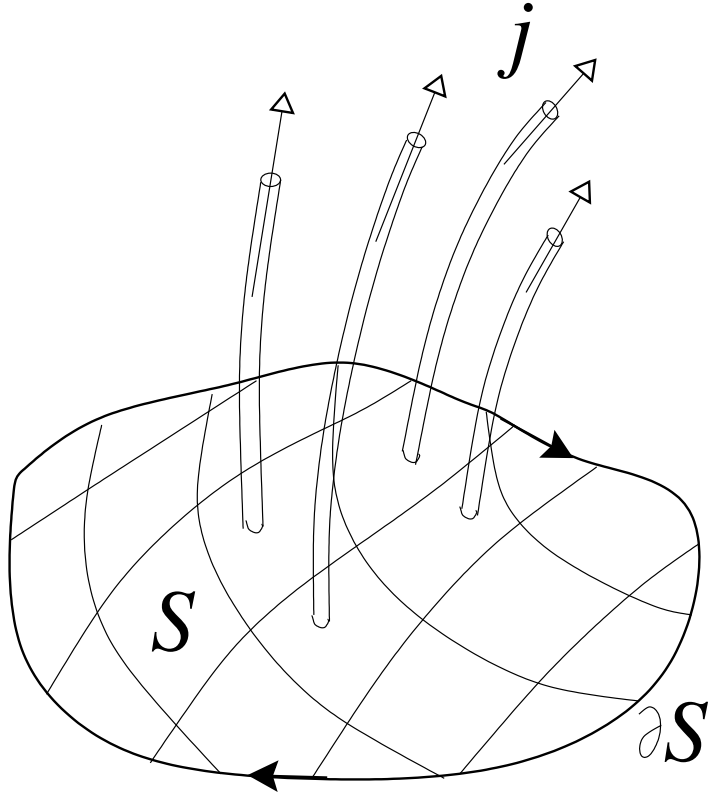} 
\caption{A surface $S$ with its boundary $\partial S$.} 
\end{center} 
\end{figure}

If we use the abbreviation $\partial_t:=\partial/\partial t$, the 
{\it global} electric charge conservation can be expressed as 
\begin{equation}
\partial_t\int\limits_V \rho + \int\limits_{\partial V}j = 0 
\qquad {\rm (Axiom \>\> 1)}\,,\label{intJ} 
\end{equation} 
where the surface integral is evaluated over the (closed and 2-dimensional) 
boundary of $V$, which we denote by $\partial V$, see Fig.2.  The change per 
time interval of the charge contained in $V$ is balanced by the flow of the 
electric current through the boundary of the domain.

The closed surface integral $\int_{\partial V}j\,$ can be transformed
into a volume integral $\int_Vd\,j$ by Stokes's theorem (\ref{Stokes}). 
Here $d$ denotes the exterior derivative that increases the rank of a 
form by one, i.e.\ $d\,j$ is a 3-form. Thus (\ref{intJ}) translates into
\begin{equation}
\int\limits_V\left({\partial_t\rho} + d\,j\right) = 0\,.\label{intJ1}
\end{equation}
Since this is valid for an arbitrary domain, we arrive at the 
{\it local form} of electric charge conservation,
\begin{equation}
d\,j + {\partial_t\rho} = 0\,.\label{Axiom1}
\end{equation}

\section{Excitations}\label{excitations}

Since the charge density $\rho$ is a 3-form, its exterior derivative
vanishes: $d\,\rho =0$. Then, by a theorem of de Rham, it follows that
$\rho$ can be derived from a ``potential'': 
\begin{equation}
d\rho = 0 \qquad\Longrightarrow\qquad \rho=d\,{\cal D}\,.\label{drho} 
\end{equation} 
In this way one finds the electric excitation 2-form ${\cal D}$. 
Its absolute dimension is $[{\cal D}] = [\rho] = q$, furthermore, for 
the components, $[{\cal D}_{ab}] = [{\cal D}]/\ell^{2} = q/\ell^{2}$.

Substituting (\ref{drho})$_2$ into charge conservation (\ref{Axiom1})
and once again using the de Rham theorem, we find another ``potential"
for the current density: 
\begin{equation}
d\left(j + {\partial_t {\cal D}}\right) = 0\qquad\Longrightarrow\qquad 
j + {\partial_t {\cal D}} = d\,{\cal H}\,.\label{jdH} 
\end{equation}
The magnetic excitation ${\cal H}$ turns out to be a 1-form, see Fig.4.  
Its dimension is $[{\cal H}]=[j]=q/t\,,\quad [{\cal H}_a]=q/(t\,\ell)$.
\begin{figure} 
\begin{center} 
\includegraphics{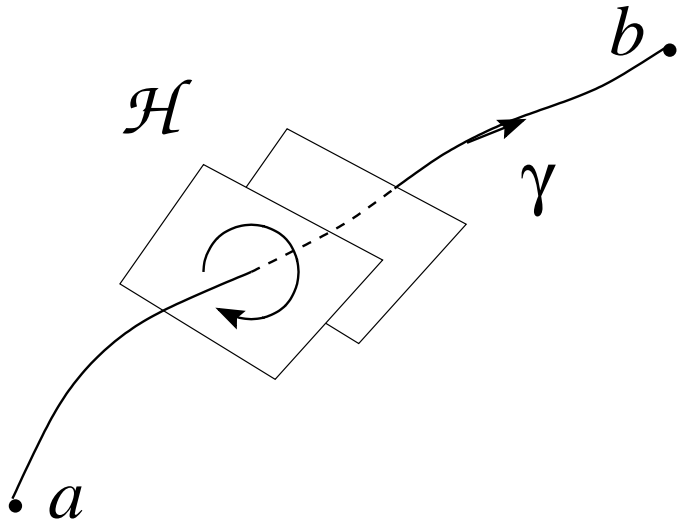} 
\caption{A line $\gamma$ with its boundary $\partial\gamma$, 
i.e., its end points $a$ and $b$.} 
\end{center} 
\end{figure} 

Consequently, the excitations $({\cal D}, {\cal H})$ are potentials of
the sources $(\rho, j)$.  All these (additive) {\em quantities} (How
much?) are described by odd differential forms.

In this way, we find the {\em inhomogeneous Maxwell} equations 
(the Gauss law and the Oersted-Amp\`ere law):
\begin{eqnarray}
  d\,{\cal D} &=& \rho\,, \label{Gauss}\\ 
  d\,{\cal H} - {\partial_t {\cal D}} &=& j\,.\label{Oersted}
\end{eqnarray}
Electric charge conservation is valid in microphysics. Therefore the
corresponding Maxwell equations (\ref{Gauss}) and (\ref{Oersted}) are
valid on the same ``microphysical'' level as the notions of charge
density $\rho$ and current density $j$. And with them the excitations
${\cal D}$ and ${\cal H}$ are microphysical quantities of the same
type likewise -- in contrast to what is stated in most textbooks.

Before we ever talked about {\em forces} on charges, charge conservation 
alone gave us the inhomogeneous Maxwell equations including the
appropriate dimensions for the excitations ${\cal D}$ and ${\cal H}$. 

Under the assumption that ${\cal D}$ vanishes inside an ideal electric
conductor, one can get rid of the indeterminacy of ${\cal D}$ which is
inherent in the definition of the excitation as a ``potential" of the
charge density, and we can measure ${\cal D}$ by means of two
identical conducting plates (``Maxwellian double plates'') which touch
each other and which are {\em separated} in the ${\cal D}$-field to be
measured. The charge on one plate is then measured. Analogously,
${\cal H}$ can be measured by the Gauss compensation method or by a
superconductor and the Meissner effect ($B=0\rightarrow {\cal H}=0$).
Accordingly, the excitations {\it do have a direct operational}
significance.

\section{Field strengths}\label{strengths}

So far, conserved charge was the notion at center stage. Now {\em
 energy} enters the scene, which opens the way for introducing the
electromagnetic field strengths. Whereas the excitations $({\cal D},
{\cal H})$ are linked to (and measured by) the charge and the current
$(\rho, j)$, the electric and magnetic field strengths are usually
introduced as forces acting on unit charges at rest or in motion,
respectively.

Let us consider a point particle with electric charge $e$ and velocity
vector $v$. The force $F$ acting on it is a 1-form since its
(1-dimensional) line integral yields a scalar, namely the energy. Thus
$F$ carries the absolute dimension of an energy or of action over
time: $[F]={\rm energy}=h/t$, where $h$ denotes the dimension of an
action.  Accordingly, the local components $F_a$ of the force $F =
F_a\,dx^a$ possess the dimension $[F_a] = h/(t\,\ell)=$ force.

In an electromagnetic field, the motion of a test particle is 
governed by the Lorentz force: 
\begin{equation}
F = e\left(E-v\rfloor B\right)\qquad{\rm (Axiom\,2)}\,.\label{Axiom2} 
\end{equation} 
The symbol $\rfloor$ denotes the interior product of a vector 
(here the velocity vector) with a $p$-form. It decreases the rank 
of a form by 1 (see the Appendix), and since $v\rfloor B$ is to be 
a 1-form, then $B$ is a 2-form. Newly introduced by (\ref{Axiom2}) 
are the electric field strength 1-form $E$ and the magnetic field 
strength 2-form $B$. They are both {\em intensities} (How strong?).

The dimension of the velocity is $[v]=1/t$. With the decomposition
$v=v^a\partial_a$, we find for its components $[v^a]=\ell/t$. Then it
is straightforward to read off from (\ref{Axiom2}) the absolute
dimension of the electric field strength $[E]=h/(t\,q)=\phi_0/t$, with
$\phi_0:=h/q$. For its components we have $[E_a]=\phi_0/(t\,\ell)$.
Analogously, for the dimension of the magnetic field strength we find
$[B]=(h/t)/(q/t)= h/q=\phi_0$ and $[B_{ab}]=\phi_0/\ell ^{2}$,
respectively. The field $B$ carries the dimension of a magnetic flux
$\phi_0$. In superconductors, magnetic flux can come in quantized flux
tubes, so-called {\em fluxoids}, underlining the importance of the
notion of magnetic flux.

The definition (\ref{Axiom2}) of the field strengths makes sense only
if the charge of the test particle is conserved. In other words, axiom
2 presupposes axiom 1 and should not be seen as a stand alone pillar
of electrodynamics.
\medskip

\section{Magnetic flux conservation}\label{fluxcons}

Taking into account the rank (as exterior forms) of the field
strengths, the only integral we can build up from $E$ and $B$,
respectively, are line integrals and surface integrals $\int_{\rm
  line} E$ and $\int_{\rm surface}B$.  Apart from a factor $t$, the
dimensions are equal. Hence, from a dimensional point of view, it
seems sensible to postulate the conservation theorem (see Fig.5), 
\begin{equation}
{\partial_t}\int\limits_S B + \int\limits_{\partial S}E = 0 
\qquad{\rm (Axiom\,3)}\,.\label{Axiom3} 
\end{equation}
Magnetic flux conservation
(\ref{Axiom3}) has to be seen as an analog of electric charge
conservation (\ref{intJ}). Magnetic flux, though, is related to a
2-dimensional surface whereas electric charge is related to a
3-dimensional volume. Thus the integration domains in the conservation
theorems (\ref{Axiom3}) and (\ref{intJ}) differ by one spatial
dimension always.
\medskip

Axiom 3 gains immediate evidence from the dynamics of an Abrikosov
flux line lattice in a superconductor. There the quantized flux lines
can be counted, they don't vanish nor are they created from nothing,
rather they move in and out crossing the boundary $\partial S$ of the
surface $S$ under consideration.
\begin{figure} 
\begin{center} 
\includegraphics{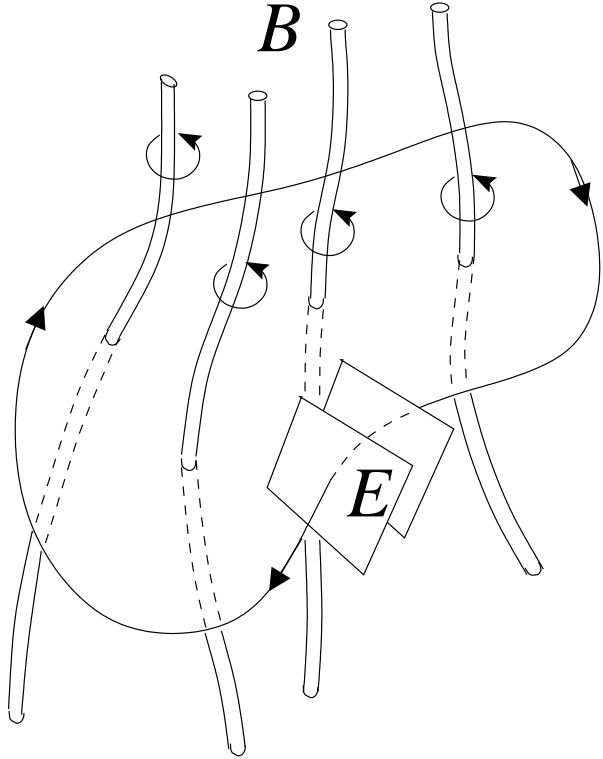} 
\caption{Faraday's induction law.} 
\end{center} 
\end{figure}

Again, by means of Stokes's theorem (\ref{Stokes}), we can transform the 
boundary integral: $\int_{\partial S}E=\int_{S}d\,E$. Taking into account 
the arbitrariness of the surface $S$, we recover Faraday's induction law
\begin{equation}
d\,E + {\partial_t B} = 0\,,\label{Farad}
\end{equation}
which is experimentally very well established. We differentiate 
Faraday's law by means of $d$ and find $\partial_t(d\,B) = 0$. Since an 
integration constant other than zero is senseless (recall the relativity 
principle), we have 
\begin{equation}
d\,B = 0\,.\label{dB0} 
\end{equation}
The {\it homogeneous Maxwell} equations (\ref{Farad}) and (\ref{dB0}) 
(Faraday's induction law and the closure of the magnetic field strength) 
nearly complete the construction of the theory.

We find the $3+3$ time evolution equations (\ref{Oersted}) and
(\ref{Farad}) for the electromagnetic field (${\cal D}, {\cal H}$; $E,
B$), i.e., for $6+6$ components. Before we can find solutions of these
equations, we have to reduce the number of variables to 6, i.e., we
have to cut them in half. Such a reduction is achieved by Axiom 4.

\section{The Maxwell-Lorentz spacetime relations}

In Axiom 4 we assume linear, isotropic, and centrosymmetric relations
between the (additive) quantities and the intensities
\cite{extensive}:
\begin{equation}
  {\cal D} = \varepsilon_0\,^\star E\qquad{\rm and}\qquad {\cal H} =
  {\frac 1 {\mu_0}}\,^\star B\qquad{\rm (Axiom\,4)}.\label{linearlaw}
\end{equation}
The proportionality coefficients $\varepsilon_0, \mu_0$ encode all the
essential information about the electric and magnetic properties of
spacetime. The Hodge star operator $\star$ is needed, since we have to
map a 1-form into a 2-form and vice versa or, more generally, a
$k$-form into a $(3-k)$-form. Then the operator $\star$ in
(\ref{linearlaw}) has the dimension of a length or its reciprocal.
Note that the Hodge star depends on the metric of our Euclidean space,
see the Appendix.  Recalling the dimensions of the excitations and the
field strengths, we find the dimensions of the electric constant
$\varepsilon_0$ and the magnetic constant $\mu_0$ as
\begin{equation}
[\varepsilon_0]={\frac{qt}{\phi_0\ell}}={\frac 1{c\,\Omega_0}}\qquad 
{\rm and}\qquad[\mu_0]={\frac{\phi_0t}{q\ell}}={\frac {\Omega_0}c}\,,
\label{epsmu0}
\end{equation}
respectively. They are also called vacuum permittivity and vacuum
permeability, see the new CODATA report \cite{CODATA}. Here we define
$\Omega_0 := \Phi_0/q = h/q^2$ and velocity $c := \ell/t$.
Dimensionwise, it is clearly visible that
\begin{equation}
\left[{\frac 1{\sqrt{\varepsilon_0\mu_0}}}\,\right] = c
\,\qquad {\rm and}\qquad 
\left[\sqrt{\frac{\mu_0}{\varepsilon_0}}\,\right]=\Omega_0\,.\label{cOmega}
\end{equation}
Obviously, the velocity $c$ and the resistance $\Omega_0$ are
constants of nature, the velocity of light $c$ being a universal one,
whereas $\Omega_0$, the characteristic impedance (or wave resistance)
of the vacuum \cite{admittance}, seemingly refers only to
electromagnetic properties of spacetime. Note that $1/\Omega_0$ plays
the role of the coupling constant of the electromagnetic field which
enters as a factor into the free field Maxwell Lagrangian.

The Maxwell equations (\ref{Gauss})-(\ref{Oersted}) and
(\ref{Farad})-(\ref{dB0}), together with the Maxwell-Lorentz spacetime
(or aether) relations (\ref{linearlaw}), constitute the foundations of
classical electrodynamics. These laws, in the classical domain,
are assumed to be of {\it universal validity}. Only if vacuum
polarization effects of quantum electrodynamics are taken care of or
hypothetical nonlocal terms should emerge from huge accelerations,
Axiom 4 can pick up corrections yielding a nonlinear law
(Heisenberg-Euler electrodynamics, see \cite{HO}) or a nonlocal law
(Volterra-Mashhoon electrodynamics, see \cite{Muench}), respectively.
In this sense, the Maxwell equations are ``more universal'' than the
Maxwell-Lorentz spacetime relations. The latter ones are not
completely untouchable.  We may consider (\ref{linearlaw}) as
constitutive relations for spacetime itself.

\section{SI-units}

The fundamental dimensions in the SI-system for mechanics and
electrodynamics are $(\ell, t, M, q/t)$, with $M$ as mass. And for
each of those a unit was defined. However, since action -- we denote
its dimension by $h$ -- is a relativistic invariant quantity and since
the electric charge is more fundamental than the electric current, we
rather choose as the basic units 
\begin{equation}
(\ell,\,t,\,h,\,q)\,,\label{lthq}
\end{equation}
see Schouten \cite{Schouten} and Post \cite{Post1}. Thus, instead
of the kilogram and the ampere, we choose joule$\times$second (or
weber$\times$coulomb) and the coulomb: 
\begin{equation}
(m, s, W\!b\!\times\!C, C).\label{mswbc} 
\end{equation}

Numerically, in the SI-system, one puts (for historical reasons) 
\begin{equation}
\mu_0=4\pi\times10^{-7}\frac{W\!b\,s}{C\,m}\qquad
{\rm (magnetic\>constant)}.\label{mu0}
\end{equation} 
Then measurements \`a la Weber-Kohlrausch yield 
\begin{equation}
\varepsilon_0=8.85\times 10^{-12}\frac{C\,s}{W\!b\,m}
\qquad{\rm (electric\>constant)}.\label{eps0}
\end{equation}

The SI-units of the electromagnetic field are collected in Table II.

\begin{center}
\noindent{\bf Table II.} SI-units of the electromagnetic field\bigskip

\noindent\begin{tabular}{|c|c|c|}
\hline 
Field     & SI-unit of field  & SI-unit of components of field \\
\hline
${\cal D}$  & $C$           &  ${C}/{m^2}$ \\
\hline
${\cal H}$  & $A={C}/{s}$ & $A/m = C/(sm)\>\>(\rightarrow{\rm oersted})$
\\
\hline
$E$         & $W\!b/s = V$    & $V/m = W\!b/(sm)$ \\
\hline
$B$         & $W\!b$          & ${W\!b}/{m^2} = {\rm tesla} = T \>\>
(\rightarrow {\rm gauss})$ \\
\hline
\end{tabular}
\end{center}
\medskip

\section{Electrodynamics in matter}

``It should be needless to remark that while from the
mathematical standpoint a constitutive equation is a postulate or a
definition, the first guide is physical experience, perhaps fortified
by experimental data." 
\hfill C.\ Truesdell and R.A.\ Toupin (1960)

\bigskip

In a great number of the texts on electrodynamics the electric and
magnetic properties of media are described following the {\it
  macroscopic averaging} scheme of Lorentz (1916). However, this
formalism has a number of serious limitations, see the relevant
criticism of Hirst \cite{Hirst}, e.g.. An appropriate modern
presentation of this subject has been given in the textbook of Kovetz
\cite{Kovetz}.

Here we follow a consistent {\it microscopic} approach to the 
electrodynamics in media, cf.\ \cite{Hirst}. The total charge or current 
density is the sum of the two contributions originating ``from the 
inside" and ``from the outside" of the medium:
\begin{equation}
  \rho = \rho^{{\rm mat}}+\rho^{{\rm ext}}\,,\qquad j = j^{{\rm
      mat}}+j^{{\rm ext}}\,.\label{total}
\end{equation}
Hereafter, the bound charge \cite{pol} in matter is denoted by {\em
  mat} and the external charge \cite{free} by {\em ext}. The same
notational scheme will also be applied to the excitations $\cal D$ and
$\cal H$.  Bound charge and bound current are inherent
characteristics of matter determined by the medium itself. They vanish
outside the medium.  In contrast, external charge and external current
in general do not vanish outside matter. They can be prepared for a
specific purpose (such as the scattering of a current of particles by
a medium or a study of the reaction of a medium in response to a
prescribed configuration of charges and currents).

We assume that the charge bound by matter fulfills the usual 
conservation law: 
\begin{equation}
d\,j^{{\rm mat}} + {\partial_t\rho^{{\rm mat}}} = 0\,.\label{Axiom1m}
\end{equation}
Taking into account (\ref{Axiom1}), this means that there is no
physical exchange (or conversion) between the bound and the external
charges. As a consequence of this assumption, we can repeat the
arguments of Sec.\ref{excitations} that will give rise to the
corresponding excitations ${\cal D}^{{\rm mat}}$ and ${\cal H}^{{\rm
    mat}}$ as ``potentials'' for the bound charge and the bound
current. The conventional names for these newly introduced excitations
are {\em polarization} $P$ and {\em magnetization} $M$, i.e.,
\begin{equation}
{\cal D}^{{\rm  mat}}\equiv -P\,,\qquad 
{\cal H}^{{\rm mat}}\equiv M\,.\label{PM}
\end{equation} 
The minus sign is conventional, see Kovetz \cite{Kovetz}. Thus, in analogy 
to the inhomogeneous Maxwell equations, we find 
\begin{equation}
- d\,P = \rho^{\rm mat}\,,\qquad d\,M + {\partial_t P} 
= j^{\rm mat}\,.\label{dP}
\end{equation}
The identifications (\ref{PM}) are only true up to an exact form.
However, if we require ${\cal D}^{\rm mat}=0$ for $E=0$ and ${\cal
  H}^{\rm mat}=0$ for $B=0$, as we will do in (\ref{susc}), uniqueness
is guaranteed.

The Maxwell equations are {\em linear} partial differential equations.
Therefore we can define
\begin{equation}
{\cal D}^{\rm ext} := {\cal D} - {\cal D}^{\rm mat} 
= {\cal D} + P\,,\qquad 
{\cal H}^{\rm ext} := {\cal H} - {\cal H}^{\rm mat} 
= {\cal H} - M\,.\label{DHe}
\end{equation}
The external excitations $({\cal D}^{\rm ext}, {\cal H}^{\rm ext})$ can 
be understood as auxiliary quantities. 
In terms of these quantities, the {\em inhomogeneous} Maxwell
equations for matter finally read:
\begin{eqnarray}
  d\,{\cal D}^{\rm ext} &=& \rho^{\rm ext}\,,\qquad {\cal D}^{\rm ext}
  = \varepsilon_0\,^\star E + P[E,B]\,,\label{dDe}\\ d\,{\cal
    H}^{\rm ext} - {\partial_t {\cal D}^{\rm ext}} &=& j^{\rm ext}\,,
  \qquad {\cal H}^{\rm ext} = {\frac 1{\mu_0}}\,^\star B -
  M[B,E]\,.\label{dHe}
\end{eqnarray}
Here the polarization $P[E,B]$ is a functional of
the electromagnetic field strengths $E$ and $B$. In general, it can 
depend also on the temperature $T$, and possibly of other thermodynamic 
variables specifying the material continuum under consideration; similar 
remarks apply to the magnetization $M[B,E]$. The system (\ref{dDe})$_1$ 
and (\ref{dHe})$_1$ looks similar to (\ref{Gauss}) and (\ref{Oersted}). 
However, the former equations refer only to the external fields and sources. 
The {\em  homogeneous} Maxwell equations (\ref{Farad}) and (\ref{dB0}) 
remain valid in their original form.

In the simplest cases, we have the linear constitutive laws
\begin{equation}
P = \varepsilon_0\,\chi_{\rm E}\,^\star E\,,\qquad
M = {\frac 1 {\mu_0}}\,\chi_{\rm B}\,^\star B\,,\label{susc}
\end{equation} 
with the electric and magnetic \cite{chiH} susceptibilities 
$(\chi_{\rm E},\chi_{\rm B})$. With material constants 
\begin{equation}
\varepsilon := \varepsilon_0\,(1 + \chi_{\rm E})\,,\qquad 
\mu := {\frac {\mu_0} {1 - \chi_{\rm B}}}\,,\label{epsmu}
\end{equation} 
one can rewrite the material laws (\ref{susc}) as 
\begin{equation}
D^{\rm ext} = \varepsilon\,^\star E\,,\qquad 
H^{\rm ext} = \frac{1}{\mu}\,^\star B\,.\label{material}
\end{equation}

For the discussion of the concrete applications of the developed
microscopic theory in modern condensed matter physics, we refer to the
review of Hirst \cite{Hirst}.

\section{Conclusion}

The Maxwell equations
\begin{eqnarray}
  d\,{\cal D} &=& \rho\,,\qquad d\,{\cal H} - {\partial_t {\cal D}} =
  j\,,\label{Minhom} \\ d\,B &=& 0\,,\qquad d\,E + {\partial_t B} = 0\,,
\label{Mhom}
\end{eqnarray}
are the cornerstones of any classical theory of electromagnetism. As
an expression of charge and flux conservation, they carry a high
degree of plausibility as well as solid experimental support. The
Maxwell equations in this form remain valid in an accelerated
reference frame and in a gravitational field likewise, without any
change.

The Maxwell-Lorentz spacetime relations
\begin{equation}
{\cal D} = \frac 1{c\,\Omega_0}\,^\star E\qquad{\rm and}\qquad 
{\cal H} = \frac {c}{\Omega_0} \,^\star B 
\end{equation} 
are necessary for developing the Maxwellian system into a predictive
physical theory. They depend, via the star operator, on the metric of
space and are, accordingly, influenced by the gravitational field.
They are valid in very good approximation, but there are a few
exceptions known (if the Casimir effect is to be described, e.g.).

For the description of matter, the sources $(\rho,j)$ and the
excitations $({\cal D}, {\cal H})$ have to be split suitably in order
to derive, from the microscopic equations (\ref{Minhom}) and
(\ref{Mhom}), appropriate macroscopic expressions.

Summing up, we can give an answer to one of the central questions
posed by Roche \cite{Roche}: The need for the different notations and
different dimensions and units for the excitation ${\cal H}$ and the
field strength $B$ (and, similarly, for $\cal D$ and $E$) is well
substantiated by the {\it very different} geometrical properties and
physical origins of these fields, see Table I and Fig.~1. Even in
vacuum, these differences do {\em not} disappear.

\bigskip

\section*{Acknowledgments} 

We are grateful to H.\ Micklitz (Cologne) for arranging this lecture.
One of the authors (FWH) would like to thank W.\ Raith
(Berlin-Bielefeld) for an extended exchange of letters on the
fundamental structure of Maxwell's theory. Moreover, he is grateful to
R.G.\ Chambers (Bristol), A.\ Kovetz (Tel Aviv), J.\ Roche (Oxford),
and S.\ Scheidl (Cologne) for most useful remarks.

\section*{Appendix: The ABC of exterior calculus}

The formalism of exterior differential forms is widely used in
different domains of mathematics and theoretical physics. In
particular, in electromagnetic theory, exterior calculus offers an
elegant and transparent framework for the introduction of the basic
notions and for the formulation of the corresponding laws. Here, we
will give a very elementary description of the objects and operations
used above.

We will confine ourselves only to the case of a {\it 3-dimensional
  space}. Let be given the set of local coordinates $x^a =\{x^1, x^2,
x^3\}$; hereafter Latin indices $a,b,\dots$ will run over $1,2,3$.
Then the vectors $e_a = \{\partial_1, \partial_2, \partial_3\}$ will
serve as a basis of the tangent vector space at every point. The
symbol $dx^a$ denotes the set of {\it linear 1-forms} dual to the
coordinate vector basis, $dx^a(e_b) = \delta^a_b$.  An arbitrary
$k$-form can be described, in local coordinates, by its components:
$\varphi = \varphi_a\,dx^a = \varphi_1\,dx^1 + \varphi_2\,dx^2 +
\varphi_3\,dx^3$ for 1-forms and $\omega = {\frac 1 2}
\,\omega_{ab}\,dx^a\wedge dx^b = \omega_{12}\,dx^1\wedge dx^2 +
\omega_{23}\,dx^2\wedge dx^3 + \omega_{31}\,dx^3\wedge dx^1$ for
2-forms.  Any 3-form has a single nontrivial component, $\eta =
\eta_{123}\, dx^1\wedge dx^2\wedge dx^3$. When $\eta$ is smoothly
defined on the whole space, it is called a {\it volume form}.
Zero-forms are just the ordinary differentiable functions.

It is often stated that the {\it exterior product} ``$\wedge$"
generalizes the vector product. However, one should be careful with
such statements, because the vector product in the standard
3-dimensional analysis is, strictly speaking, a superposition of the
wedge product {\it and} of the Hodge duality operator. Thus, the
vector product necessarily involves the metric on the manifold. In
contrast, the exterior product is a {\it pre}-metric operation,
although it resembles the vector product. For example, the exterior
product of the two 1-forms $\omega$ and $\varphi$ with the components
$\omega_a$ and $\varphi_a$ yields a 2-form $\omega \wedge\varphi$ with
the local components $\{(\omega_2\varphi_3 - \omega_3 \varphi_2),
(\omega_3\varphi_1 - \omega_1\varphi_3), (\omega_1\varphi_2 -
\omega_2\varphi_1)\}$.

The {\it exterior differential} $d$ increases the rank of a form by 1.
It is most easily described in local coordinates, see Table III.
Thus, $d$ naturally generalizes the ``grad" operator which leads from
a scalar to a vector and, at the same time, it represents a {\it
  pre}-metric extension of the ``curl" operator. The exterior
differential is a nilpotent operator, i.e., $dd = 0$.

\begin{center}
\noindent{\bf Table III.} Operators acting on an exterior form\bigskip

\noindent\begin{tabular}{|c|c|}
\hline 
& $k$-form $\omega = {\frac 1 {k!}}\,\omega_{a_1\dots
a_k}\,dx^{a_1}\wedge\dots\wedge dx^{a_k}$, with $k = 0,1,2,3$\\
\hline
{\ }$d${\ } & $(k+1)$-form $d\omega = {\frac 1 {(k+1)!}}\,\left(\partial_{[a_1}
\omega_{a_1\dots a_{k+1}]}\right)\,dx^{a_1}\wedge\dots\wedge
dx^{a_{k+1}}$\\
\hline
{\ }$\rfloor${\ } & $(k-1)$-form $v\rfloor\omega = {\frac 1 {(k-1)!}}\,v^a
\omega_{aa_1\dots a_{k-1}}\,dx^{a_1}\wedge\dots\wedge dx^{a_{k-1}}$\\
\hline
{\ }${}^\star${\ } & $(3-k)$-form $^\star\omega = {\frac 1 {k!}}
\,\omega_{a_1\dots a_k}\,g^{a_1b_1}\dots g^{a_kb_k}\,e_{b_k}
\rfloor\dots\rfloor e_{b_1}\rfloor\eta$ \\
\hline 
\end{tabular}
\end{center}
\medskip

Complementary to $d$, one can define an operation which decreases the
rank of a form by 1. This is the {\it interior product} of a vector
with a $k$-form.  Given the vector $v$ with the components $v^a$, the
interior product with the coframe 1-form yields $v\rfloor dx^a = v^a$,
which is a sort of a projection along $v$. By linearity, the interior
product of $v$ with a $k$-form is defined as described in Table III.

The {\it Hodge dual operator} ${}^\star$ maps $k$-forms into
$(3-k)$-forms.  Its introduction necessarily requires the {\it metric}
which assigns a real number $g(u,v)=g(v,u)$ to every two vectors $u$
and $v$. In local coordinates, the components of the metric tensor are
determined as the values of the scalar product of the basis vectors,
$g_{ab}:= g(e_a,e_b)$.  This matrix is {\it positive definite}. The
metric introduces a natural volume 3-form $\eta := \sqrt{\det
  g_{ab}}\,dx^1\wedge dx^2\wedge dx^3$ which underlies the definition
of the Hodge operator ${}^\star$. The general expression is displayed
in Table III. Explicitly the Hodge dual of the coframe 1-form reads,
for example: $^\star dx^a = \sqrt{\det g_{ab}}\, (g^{a1}\,dx^2\wedge
dx^3 + g^{a2}\,dx^3\wedge dx^1 + g^{a3}\,dx^1\wedge dx^2)$, where
$g^{ab}$ is inverse to $g_{ab}$.

The notions of the {\it odd} and {\it even} exterior forms are closely
related to the orientation of the manifold. In simple terms, these two
types of forms are distinguished by their different behavior with
respect to a reflection (i.e., a change of orientation): an even (odd)
form does not change (changes) sign under a reflection transformation.
These properties of odd and even forms are crucial in the integration
theory, see, e.g., \cite{Ted}.

For a $k$-form an {\it integral} over a $k$-dimensional subspace is
defined. For example, a $1$-form can be integrated over a curve, a
$2$-form over a 2-surface, and a volume 3-form over the whole
3-dimensional space.  We will not enter into the details here,
limiting ourselves to the formulation of Stokes's theorem which
occupies a central place in integration theory:
\begin{equation}
\int\limits_{\partial C}\omega=\int\limits_{C}d\,\omega.\label{Stokes}
\end{equation}
Here $\omega$ is an arbitrary $k$-form, and $C$ is an arbitrary 
$(k+1)$-dimensional (hyper)surface with the boundary $\partial C$.

For a deeper and a more rigorous introduction into exterior calculus,
see, e.g., \cite{Bamberg,Ted}.


\begin{thebibliography}{10}

\bibitem{book} F.W.~Hehl and Yu.N.~Obukhov, {\it Electric Charge and
    Magnetic Flux: On the Structure of Classical Electrodynamics.}
  Tex-script of 230 pages. March 1999 (unpublished).

\bibitem{Zirnbauer} M.R.\ Zirnbauer, {\it Elektrodynamik.} Tex-script
  of a course in Theoretical Physics II (in German), July 1998
  (Springer, Berlin, to be published).

\bibitem{OH} Yu.N.\ Obukhov and F.W.\ Hehl, {\it Space-time metric from 
linear electrodynamics}, {\sl Phys. Lett.} {\bf B458} (1999) 466-470.

\bibitem{HO} F.W.~Hehl and Yu.N.~Obukhov, {\it How does the
    electromagnetic field couple to gravity, in particular to metric,
    nonmetricity, torsion, and curvature?} Preprint IASSNS-HEP-99-116,
  Institute for Advanced Study, Princeton, see also {\tt
    http://arXiv.org/ abs/gr-qc/0001010}.

\bibitem{Roche} J.J. Roche, {\it B and H, the intensity vectors of
    magnetism: A new approach to resolving a century-old controversy},
 {\sl Am. J. Phys.} {\bf 68} (2000) 438-449.

\bibitem{Bamberg} P. Bamberg and S. Sternberg, {\it A Course in
    Mathematics for Students of Physics.} Vol.2 (Cambridge University
  Press, Cambridge 1990).

\bibitem{Bopp} F.\ Bopp, {\it Prinzipien der Elektrodynamik}, 
{\sl Z. Physik} {\bf 169} (1962) 45-52. 

\bibitem{Ingarden} R. Ingarden and A. Jamio{\l}kowski, {\it Classical
    Electrodynamics} (Elsevier, Amsterdam 1985).

\bibitem{Kovetz} A.\ Kovetz, {\it Electromagnetic Theory.} Oxford
  University Press, Oxford, UK (2000).

\bibitem{Post1} E.J. Post, {\it Formal Structure of Electromagnetics}
  -- General Covariance and Electromagnetics. North Holland, Amsterdam
  (1962) and Dover, Mineola, New York (1997).

\bibitem{Sommerfeld} {A.\ Sommerfeld}, {\it Elektrodynamik.}
  Vorlesungen \"uber Theoretische Physik, Band 3. Dietrisch'sche
  Verlagsbuchhandlung, Wiesbaden (1948).

\bibitem{Truesdell} C.\ Truesdell and R.A.\ Toupin, {\it The Classical
  Field Theories}. In {\sl Handbuch der Physik,} Vol.\ III/1, S.
  Fl{\"u}gge ed.. Springer, Berlin (1960) pp. 226-793.

\bibitem{Chambers} R.G.\ Chambers, {\em Units --- $B,H,D$, and $E$}, {\sl
    Am. J. Phys.} {\bf 67} (1999) 468-469.

\bibitem{Rochebook} J.J.\ Roche, {\it The Mathematics of Measurement} --
    A Critical History. The Athlone Press, London (1998).

\bibitem{Fitch} V.L.\ Fitch, {\it The far side
      of sanity}, {\sl Am. J. Phys.} {\bf 67} (1999) 467.

\bibitem{Ted} T.\ Frankel, {\it The Geometry of Physics} -- An
Introduction. Cambridge University Press, Cambridge (1997). Now also
available as paperback.

\bibitem{Schouten} J. A. Schouten, {\it Tensor Analysis for Physicists}, 
2nd ed.. Clarendon Press, Oxford (1954) and Dover, Mineola, New York (1989).

\bibitem{Raith} W.\ Raith, {\it Bergmann-Schaefer, Lehrbuch der
Experimentalphysik, Vol.2, Elektromagnetismus}, 8th ed.. de
Gruyter, Berlin (1999).

\bibitem{Dehmelt} H. Dehmelt, {\it Experiments with an isolated
    subatomic particle at rest}, {\sl Rev. Mod. Phys.} {\bf 62} (1990)
  525-530.

\bibitem{Paul} W.\ Paul, {\it Electromagnetic traps for charged and
    neutral particles}, {\sl Rev. Mod. Phys.} {\bf 62} (1990) 531-540.

\bibitem{Devoret1} M.H.\ Devoret, D.\ Esteve, and C.\ Urbina, {\it
    Single-electron transfer in metallic nanostructures}, {\sl Nature}
  {\bf 360} (1992) 547-552.

\bibitem{Devoret2} P.\ Lafarge, P.\ Joyez, D.\ Esteve, C.\ Urbina, and
  M.H.\ Devoret, {\it Two-electron quantization of the charge on a
    superconductor}, {\sl Nature} {\bf 365} (1993) 422-424.

\bibitem{extensive}{Alternatively, between the extensive and the
    intensive quantities. In electrodynamics, the distinction between
    these two types of quantities goes back to Mie. He also suggested
    the name excitation (in German: Erregung) for ${\cal D}$ and
    ${\cal H}$, see G.~Mie: {\em Lehrbuch der Elektrizit\"at und des
      Magnetismus.} 3rd ed..  Enke, Stuttgart (1948), see also
    \cite{Sommerfeld} and \cite{Bamberg}}.

\bibitem{CODATA} P.J.\ Mohr and B.N.\ Taylor, {\it CODATA recommended
    values of the fundamental physical constants: 1998}, {\sl Rev.
    Mod. Phys.} {\bf 72} (2000) 351-495.

\bibitem{admittance} {$1/\Omega_0$ is also sometimes called the
      admittance of free space, see, for instance, \cite{Post1} p.\ 
      184.}

\bibitem{Muench} U.\ Muench, F.W.\ Hehl, and B.\ Mashhoon, {\it
      Acceleration-induced nonlocal electrodynamics in Minkowski
      spacetime}, {\sl Phys. Lett.} {\bf A271} (2000) 8-15.

\bibitem{Hirst} L.L.\ Hirst, {\it The microscopic magnetization: 
concept and application}, {\sl Rev. Mod. Phys.} {\bf 69} (1997) 607-627.

\bibitem{pol}{Also called polarization charge.}

\bibitem{free}{Also called free, true, or real charge.}

\bibitem{chiH}{In older texts, the magnetization $M$ was usually
    expressed in terms of H, namely $M = \chi_{\rm H}\,H$.  For reasons
    of relativistic invariance, this is inappropriate, provided we
    start with $P = \varepsilon_0\,\chi_{\rm E}\,^\star E$, as we do in
    (\ref{susc})$_2$. Note that $\mu = \mu_0 (1 + \chi_{\rm H}).$}

\end{thebibliography}
\end{document}